\documentclass[pre,twocolumn,showpacs,preprintnumbers,amsmath,amssymb]{revtex4}
\usepackage{graphicx}
\usepackage{dcolumn}
\usepackage{bm}

\def \be{\begin{equation}}
\def \ee{\end{equation}}
\def \bea{\begin{eqnarray}}
\def \eea{\end{eqnarray}}
\def \ba{\begin{array}}
\def \ea{\end{array}}
\def \non{\nonumber}
\begin{document}


\title{Fluctuation-dissipation relations in driven dissipative systems}
\author{Yair Shokef}
\thanks{Formerly Srebro}
\author{Guy Bunin}
\author{Dov Levine}
\affiliation{Department of Physics, Technion, Haifa 32000, Israel}
\date{\today}

\begin{abstract}

Exact theoretical results for the violation of time dependent fluctuation-dissipation relations in driven dissipative systems are presented. The ratio of correlation to delayed response in the stochastic model introduced in [Phys. Rev. Lett. 93, 240601 (2004)] is shown to depend on measurement time. The fluctuation temperature defined by this ratio differs both from the temperature of the environment performing the driving, and from other effective temperatures of the system, such as the average energy (or ``granular temperature''). General explanations are given for the time independence of fluctuation temperature for simple measurements or long measurement times.

\end{abstract}


\maketitle


The fluctuation-dissipation relation (FDR) provides a fundamental connection between two small deviations from thermodynamic equilibrium - the autocorrelation function of an observable, and the response function  of that observable to changes in its conjugate field. In equilibrium, there are two aspects to the FDR: (i) these functions have the same spatial and temporal (or frequency) dependence, and (ii) for every observable their ratio equals the system's temperature. 

Away from equilibrium, there is no such theorem connecting correlation and response. This having been said, there is much theoretical \cite{t_fd,more_t_fd,glass_observables,garzo,srebro_levine,t_fd_glass,t_fd_foam,compaction_models,t_fd_gran,fd_ohern_prl_2004,jap_tfd_vs_t,baldassarri_2005} and experimental \cite{t_fd_exp,danna_2003} interest as to what the relation between these quantities is. In particular, one may ask whether either aspect (i) or (ii) holds for small deviations from non-equilibrium steady states (NESS). Aspect (ii) has been tested extensively for low frequency measurements in various systems, and the effective temperature $T_F$, defined by the ratio of correlation to response, clearly differs from the environment temperature \cite{t_fd,more_t_fd,t_fd_exp}. On the other hand, numerical simulations indicate that $T_F$ does coincide \cite{footnote_coincidence} in some cases with other effective temperatures defined for glasses \cite{t_fd_glass}, sheared dissipative foam \cite{t_fd_foam}, dense granular packings \cite{compaction_models} and driven granular gases \cite{t_fd_gran}.

Regarding aspect (i), different frequency dependence of the correlation and response functions has been observed in simulations of glasses \cite{t_fd}, sheared foam \cite{fd_ohern_prl_2004} and models of driven systems \cite{jap_tfd_vs_t}, as well as in bacterial bath experiments \cite{lubensky}. In all these cases, fast modes thermalize, while slower modes manifest a higher (usually frequency-independent) temperature reflecting driving or system history (e.g., temperature quench). Experiments on vibrated granular systems \cite{danna_2003} have shown a weak dependence of $T_F$ on measurement frequency, which may not obviously be attributed to thermalization of fast modes with the environment. 

This work aims at understanding the relation between time- (or frequency-) dependent correlations and response in NESS. To our knowledge, these have not previously been calculated exactly for any driven dissipative system. Here we calculate the autocorrelation and response functions in the context of a simple stochastic model of a dissipative system. The model introduced in \cite{srebro_levine} is constructed about the essential features of any driven dissipative system: energy loss upon interaction between particles (or modes) and energy input through an external driving mechanism. As such, we have reason to hope that our results will be relevant to a broad class of driven dissipative systems \cite{footnote_ener_dist}. 

Our main results are: (1) There are certain observables for which correlation and response have identical temporal dependence, however, (2) For a general measurement, they have different temporal dependences. This notwithstanding, (3) For asymptotically long times, the time dependence of correlation and response coincides. Spatial FDR in our model are investigated in \cite{dana}.


The model consists of $N$ particles, each with two degrees of freedom (DOF), one ``kinetic'', $e_{i}$, and one ``internal'', $x_{i}$. The energy of particle $i$ is $e_{i} -  x_{i}F_{i}$, where $F_{i}$ is an external field on particle $i$, which may depend on time. The $\{ x_{i} \}$ may be thought of as positions or any other coordinates used to measure FDR. In our notation we emphasize the difference between \emph{microscopic} DOF's $\{ e_i \}$ and $\{ x_i \}$ and \emph{macroscopic} quantities $E \equiv \sum_{i=1}^N e_i$ and $X \equiv \sum_{i=1}^N x_i$. The system dissipates energy through interactions  and is maintained in a steady state by being coupled to a thermal bath.

In every interaction two DOF's are randomly chosen and their energy is stochastically redistributed between them. In interactions between two ``kinetic'' DOF's ($e-e$) there is dissipation: only a fraction $0 \leq \alpha \leq 1$ ($\alpha$ is like a restitution coefficient) of the energy is conserved and the remaining is dissipated out of the system. Interactions between a ``kinetic'' and an ``internal'' DOF ($e-x$), and between a ``kinetic'' DOF and a DOF from the bath ($e-e_{B}$) conserve energy. When all interactions are elastic $( \alpha = 1 )$, detailed balance holds and the system reaches thermodynamic equilibrium with the bath.
Rates of interactions are determined by the per-particle interaction rate $\Gamma$, and the dimensionless coupling strengths $0 \leq ( f , h ) \leq 1$ of the ``kinetic'' DOF's to the bath and to the ``internal'' DOF's, respectively. 

The resulting stochastic equation of motion for particle $i$ reads,
\begin{subequations}\label{eq:dyn_rule_fd}
\bea e_i(t+dt)= \left\{
\begin{array}{cc}
  \underline{value:} & \underline{probability:} \\
  e_i(t) & 1- \Gamma dt \\
  z \alpha [e_i(t)+e_k(t)]  & (1-f) \Gamma dt \\
  z [e_i(t)+e_{B}]  & f (1-h) \Gamma dt \\
  z [e_i(t)-x_j(t) F_j(t)]  & f h \Gamma dt
\end{array} \right.\label{eq:dyn_rule_fd_e}
\eea
\bea x_i(t+dt)= \left\{
\begin{array}{cc}
  \underline{value:} & \underline{probability:} \\
  x_i(t) & 1- f h \Gamma dt \\
  z \left[x_i(t)- e_j(t) / F_i(t) \right]  & f h \Gamma dt
\end{array} \right.\label{eq:dyn_rule_fd_x}
\eea
\end{subequations}
where:  $j,k \in \{ 1 , ..., N \}$ are indices of particles with which particle $i$ may interact, chosen randomly at every interaction (we do not allow $e_i-e_i$ interactions, therefore $k \neq i$); $z \in [0,1]$ is the fraction of repartitioned energy given to particle $i$ in the interaction, chosen randomly from a uniform distribution at every interaction; $e_{B}$ is the energy of a bath DOF chosen randomly at every interaction from the equilibrium distribution $p_B(e_{B}) = T_{B}^{-1} e^{-e_{B}/T_{B}}$ at the environment temperature $T_B$. 


In the steady state we define \emph{1-particle} and \emph{N-particle} autocorrelation functions, $c(t) \equiv \langle x_i(t) x_i(0) \rangle - \langle x_i \rangle ^2$ and $C(t) \equiv \langle X(t) X(0) \rangle - \langle X \rangle ^2$, respectively, and corresponding delayed response functions $r(t) \equiv \partial \langle x_i(t) \rangle / \partial F_i(0)$ and $R(t) \equiv \partial \langle X(t) \rangle / \partial F(0)$. The responses $r(t)$ and $R(t)$ are to a sudden change at time $t=0$ of the field on the single particle $i$ or of the uniform field on all particles, respectively. $\langle \: \rangle$ denotes the steady state ensemble average over possible states at the beginning of the measurement ($t=0$). In equilibrium ($\alpha=1$ or $f=1$), $c(t)=r(t)T_{B}$ for any $x_i$, and $C(t)=R(t)T_{B}$ for $X$ \cite{footnote_fdt}; we will be concerned with the non-equilibrium cases.


For 1-particle FDR we consider a system evolving from a given state at $t=0$ until a later time $t$. We average Eq. (\ref{eq:dyn_rule_fd_x}) over the stochasticity in the dynamics for this initial state, and denote this averaging by an overline,
\bea
\frac{2}{\Gamma} \frac{d  \overline{x_i(t)} }{dt} = -fh \left( \overline{x_i(t)} + \frac {\overline{E(t)}}{N F_i(t)} \right) \label{eq:dxi_dt} .
\eea
By the central limit theorem, for large systems ($N \gg 1$) the relative fluctuation $(x_i - \langle x_i \rangle) / \langle x_i \rangle$ of $x_i$ is much larger than the relative fluctuation $(E - \langle E \rangle) / \langle E \rangle$ of $E$, thus the steady state average $\langle E \rangle$ may be substituted for the instantaneous value $\overline{E(t)}$ in Eq. (\ref{eq:dxi_dt}). For a constant field, noting that $\langle e \rangle \equiv \langle E \rangle / N = - \langle x_i \rangle F_i$ \cite{footnote_F_0}, this yields,
\bea
\overline{x_i(t)} = \langle x_i \rangle + (x_i(0) - \langle x_i \rangle) e^{- \gamma t} \label{eq:xit},
\eea
with $\gamma \equiv f h \Gamma / 2$. The correlation is obtained by multiplying Eq. (\ref{eq:xit}) by $x_i(0)$ and averaging over the steady state distribution for all possible initial ($t=0$) states of the system, which yields $c(t) = (\langle x_i^2 \rangle - \langle x_i \rangle ^2) e^{- \gamma t}$ \cite{2nd_moms}.

The 1-particle response is obtained by maintaining the system in a steady state with some external field, then, at time $t=0$, changing the field $F_i$ acting on particle $i$ from $F_0$ to $F$. We then follow $\overline{x_i(t)}$ and average over the system's steady state distribution at time $t=0$ to yield $\langle x_i(t) \rangle$. The steady state solution of Eq. (\ref{eq:dxi_dt}) yields $- \langle x_i(0) \rangle F_0 = - \langle x_i \rangle F = \langle e \rangle$, thus $\langle x_i(t) \rangle = - \langle e \rangle [F^{-1} + (F_0^{-1}-F^{-1}) e^{- \gamma t} ]$. After differentiating with respect to $F_0$ and taking the limit $F_0 \rightarrow F$, we get $r(t) = \langle e \rangle F^{-2} e^{- \gamma t}$. 

We thus see that the 1-particle correlation and response have the same temporal dependence and aspect (i) of the FDR exactly holds, with the time-independent fluctuation temperature \cite{2nd_moms},
\bea
T_{F}^{1} \equiv \frac{c(t)}{r(t)} = T_G \frac{\langle x_i^2 \rangle - \langle x \rangle ^2}{\langle x \rangle ^2} \label{eq:tfd_mic}.
\eea
Aspect (ii), on the other hand, is violated \cite{srebro_levine}, since $T_F^1$ generally differs both from the environment temperature $T_B$ and from the granular temperature, defined as the average energy per DOF: $T_G \equiv \langle e \rangle$. Only in the equilibrium limits ($\alpha \rightarrow 1$, $f \rightarrow 1$) is the distribution of $x_i$ the exponential Boltzmann distribution and $T_F^1=T_G=T_B$. 


The FDR is only proven for equilibrium, so one might expect the correlation and response to depend differently on time in a NESS. However, the correlation represents the system's return to steady state after deviating from it due to spontaneous fluctuations, and the response represents its return to steady state after being moved away from it by some external force. Both are governed by the same physical processes and thus generally posses the same time scales. In simple cases (as the 1-particle measurement solved above) there is only one time scale and the correlation and response are not rich enough to have different time dependences. In order to observe violations of aspect (i) in the FDR we now consider a measurement on a macroscopic quantity. This possesses two time scales - just enough to exhibit a different temporal dependence for the correlation and response.


We assume uniform $F$ and sum Eq. (\ref{eq:dxi_dt}) over particles,
\bea
\frac{2}{\Gamma} \frac{d \overline{X(t)} }{dt} = -fh \left( \overline{X(t)} + \frac {\overline{E(t)}}{F(t)} \right)
\label{eq:dX_dt} .
\eea
Now, fluctuations in $E$ are not negligible compared to fluctuations in $X$, so Eq. (\ref{eq:dX_dt}) must be solved in conjunction with the equation similarly derived from Eq. (\ref{eq:dyn_rule_fd_e}), 
\bea
\frac{2}{\Gamma} \frac{d \overline{E(t)} }{dt} = -fh \overline{X(t)} F(t) - A_1 \overline{E(t)} + f (1-h) N T_B \label{eq:dE_dt} ,
\eea
where $A_n \equiv n + (1-f)(1-2 \alpha^n)$. The average evolution of the system from a state with $X(0)$ and $E(0)$ is hence given by the simultaneous solution of Eqs. (\ref{eq:dX_dt}) and (\ref{eq:dE_dt}):
\bea
&\overline{X(t)}& = \langle X \rangle \non \\ &+& \sum_{\ell=1}^2 \left[ a_{\ell} \left( X(0)-\langle X \rangle \right) + b_{\ell} \frac{E(0) - \langle E \rangle}{F} \right] e^{-\gamma_{\ell} t} \label{eq:xbar_vs_t}
\eea
where we denote $k \equiv \left(A_1^2 - 2A_1fh + 5f^2h^2 \right) ^ {1/2}$, $a_1 \equiv (1 - A_1 + fh)/2k$, $a_2 \equiv (1 + A_1 - fh)/2k$, $b_1 \equiv  fh/k$, $b_2 \equiv - fh/k$, $\gamma_1 \equiv (A_1+fh + k) \Gamma / 4$ and $\gamma_2 \equiv (A_1+fh - k) \Gamma / 4$,
and the steady state values are given by $ - \langle X \rangle F = \langle E \rangle = f(1-h) N T_B/(A_1-fh)$.

To compute the correlation we multiply $\overline{X(t)}$ by $X(0)$ and average over the steady state distribution of the initial states (that is, over $X(0)$ and $E(0)$) \cite{2nd_moms}:
\bea 
C(t) &=&  \sum_{\ell=1}^2 \left[ a_{\ell} \left( \langle X^2 \rangle - \langle X \rangle ^2 \right) \right. \non \\ &+& \left. b_{\ell}  \frac{\left\langle X E \right\rangle  - \langle X \rangle \langle E \rangle }{F}  \right] e^{-\gamma_{\ell} t} \label{eq:C_vs_t}.
\eea

For the response we change the field from $F_0$ to $F$ at $t=0$, and after averaging Eq. (\ref{eq:xbar_vs_t}) over the initial states taken from the steady state corresponding to the field $F_0$, differentiating with respect to $F_0$ and taking the limit $F_0 \rightarrow F$ we have,
\bea
R(t) = \frac{\langle E \rangle}{F^2} \sum_{\ell=1}^2 a_{\ell} e^{-\gamma_{\ell} t} \label{eq:R_vs_t}.
\eea

\begin{figure}[tb]
\includegraphics[width=8cm]{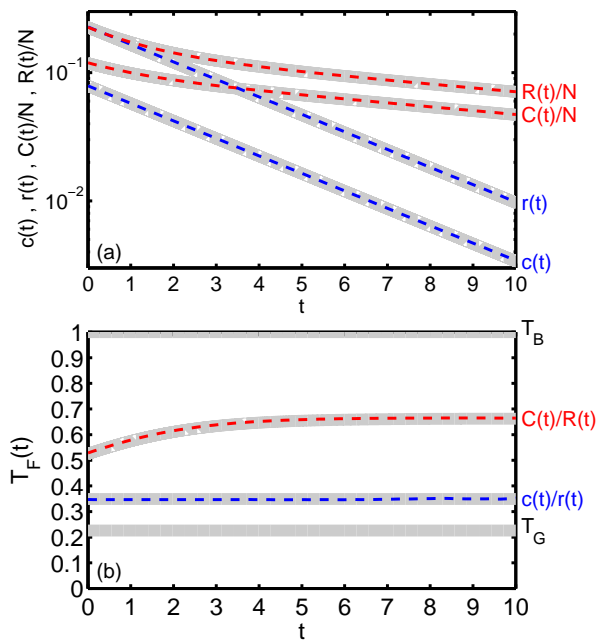}
\caption{\label{fig:tfd_vs_t} (color online) (a) Correlation and response functions and (b) resulting fluctuation temperatures vs. measurement time for the model with $\alpha=0.6$, $f=0.7$ and $h=0.9$. Graphs are scaled to $T_B=1$, $F=1$ and $\Gamma=1$. The resulting decay rates are $\gamma=0.32$ , $\gamma_1=0.72$ and $\gamma_2=0.07$. Numerical simulations with $N=10^3$ (dashed lines) agree with theoretical results for $N \gg 1$ (thick gray lines). The bath temperature $T_B$ and the granular temperature $T_G$ are given for reference.}
\end{figure}

In this N-particle measurement $C(t)$ and $R(t)$ share the rates $\gamma_1$ and $\gamma_2$, however with different prefactors, leading to violation of aspect (i) in the FDR \cite{glass_footnote}. If one insists on defining a fluctuation temperature $T_{F}^{N}(t) \equiv C(t)/R(t)$, it will depend on the measurement time, as can be seen in Fig. \ref{fig:tfd_vs_t},
which also demonstrates that the 1-particle and N-particle measurements yield fluctuation temperatures differing both one from the other as well as from the granular temperature $T_G$ and bath temperature $T_B$.


As the correlation and response generally share the same time scales, for measurement times longer than the maximal time scale of the system ($1/\gamma_2$ for the N-particle measurement in our model) this time scale dominates both correlation and response and aspect (i) of the FDR asymptotically holds for long times (see Fig. \ref{fig:tfd_vs_t}). It is interesting to speculate that this behavior may hold for general dissipative systems.


Although the FDR is generally not valid in NESS, it is hard to observe its violations in simulations or experiments on dissipative systems. One reason for this is that dissipative systems often have nearly Boltzmann distributions with some effective temperature, and deviations between different effective temperatures of the system (like $T_{F}$ and $T_G$) are too small to be observed \cite{garzo,srebro_levine}. 

Another reason is that some measurements on some systems do not exhibit at all the FDR violations presented above. To demonstrate this we now show that in the inelastic Maxwell model \cite{maxwell_model} the frequency dependent Kubo formula \cite{kubo_stat_phys_2} holds exactly with an effective temperature coinciding with $T_G$. That is, the velocity autocorrelation $D(\omega) \equiv \int_0^{\infty} \langle v(0) v(t) \rangle e^{-i \omega t} dt$ relates to the mobility $\mu (\omega)$ by $D(\omega) = \mu(\omega) T_{K}$ with $T_K=T_G$ independent of frequency \cite{baldassarri_footnote}. We shall first show that for our model $T_K$ is frequency-independent and equals $T_F^1$, which generally differs from $T_G$, and subsequently show that for the Maxwell model $T_K=T_G$.


To make contact with the standard Kubo relations we will imagine the DOF's $\{ x_i \}$ in our model as positions; thus the autocorrelation of the single particle ``velocity'' $v_i \equiv dx_i/dt$ is obtained by twice differentiating $c(t)$ calculated above. The relation $c(t)=r(t) T_F^1$ with $T_F^1$ independent of $t$, as obtained for the 1-particle FDR solved above, may be twice differentiated with respect to time and transformed to frequency domain. The Kubo formula then immediately follows with $T_K=T_{F}^1$. We have also verified this general result by explicitly calculating $D(\omega)$ and $\mu(\omega)$ in our model and obtained $D(\omega) = \mu(\omega) T_{F}^1 = (\langle x_i^2 \rangle - \langle x_i \rangle ^2) / (\gamma + i \omega)$.


Our method may now be applied to the $1D$ inelastic Maxwell model (the $2D$ and $3D$ versions follow trivially): During a finite time $d t > 0$, short compared to the collision rate $\Gamma$ and to the coupling rate $\lambda$ to the thermostat, the effect of the fluctuating force in the Langevin dynamics each particle undergoes is proportional to $\sqrt {d t}$ (see e.g. \cite{williams_mackintosh}). Thus the velocity of particle $i$ evolves as,
\bea v_i(t + d t)= \left\{
\begin{array}{cc}
  \underline{value:} & \underline{probability:} \\
  ( 1 - \lambda d t ) v_i(t) + \psi_i(t) \sqrt{d t} & 1- \Gamma d t \\
  \frac{1-\alpha}{2} v_i(t) + \frac{1+\alpha}{2} v_j(t) & \Gamma d t \\
\end{array} \right.\label{eq:dyn_rule_mm_v}
\eea
For stochastic thermostating, $\psi_i(t)$ is an uncorrelated random force with $\langle \psi^2 \rangle =  2 \lambda T_B$, while for Gaussian thermostating, $\psi_i(t) = 0$ and $\lambda = (\alpha^2-1) \Gamma / 4$. Averaging Eq. (\ref{eq:dyn_rule_mm_v}) yields $\langle v(0) v(t) \rangle = \langle v^2 \rangle e^{-\kappa t}$ with $\kappa \equiv \lambda + (1+\alpha) \Gamma /2$, thus $D(\omega) = \langle v^2 \rangle  / (\kappa + i \omega)$. In the mobility measurement, a periodic acceleration $\xi(t) = \xi_0 e^{i \omega t}$ is added to a single particle, and from Eq. (\ref{eq:dyn_rule_mm_v}) we have
\bea
\frac{d\overline{v_i(t)}}{dt} = - \kappa \overline{v_i(t)} + \xi_0 e^{i \omega t} .
\eea
This has the steady solution $\langle v_i(t) \rangle = \xi_0 e^{i \omega t} / (\kappa + i \omega)$, thus $\mu(\omega) = 1 / ( \kappa + i\omega)$, and the Kubo formula holds with $T_K=T_G$. 

This difference between our model and the Maxwell model derives from the fact that $T_F$ and $T_K$ measure fluctuations, expressed in second moments of DOF's. In the Maxwell model $T_G$ is defined from $\langle v^2 \rangle$, which is the second moment of a DOF, while in our model $T_G \equiv \langle e \rangle$ is defined from the first moment of a DOF. Only in equilibrium do all moments of the exponential Boltzmann distribution yield a single energy scale and these measurements in both models give the actual temperature.


To our knowledge, we have presented the first exact results for time dependent violations of FDR in driven dissipative systems by demonstrating that correlation and response functions have different temporal behaviors. This violation, like differences between different effective temperatures of a system, may be small, requiring sensitivity in simulations and experiments. We identified situations relevant to general driven dissipative systems where the correlation and response have the same time dependence: In sufficiently simple measurements the correlation and response share a single time scale and so have the same dependence on time, thus time-dependent FDR violations may be observed only in systems or measurements exhibiting multiple time scales. Finally, since the correlation and response typically share the same time scales, their temporal behavior for measurements with long waiting times asymptotically coincides.


We thank Naama Brenner, J. Robert Dorfman, Sam F. Edwards, Dmitri Grinev, Fred MacKintosh, Jorge Kurchan, Dana Levanony and Yael Roichman for helpful discussions. DL acknowledges support from grants 88/02 and 660/05 of the Israel Science Foundation and the Fund for the Promotion of Research at the Technion.




\begin{thebibliography}{10}
\bibitem{footnote_coincidence}
Dependence of $T_F$ on observable was found in glasses \cite{glass_observables}, while kinetic theory approximations for granular gases \cite{garzo} as well as exact solutions of a model system \cite{srebro_levine} showed $T_F$ differs from the granular temperature.
\bibitem{footnote_ener_dist}
For example, our model qualitatively reproduces the energy distribution of driven granular gases \cite{srebro_levine}.
\bibitem{footnote_F_0}
$\langle x \rangle$, $c(t)$, $r(t)$, $C(t)$ and $R(t)$ all diverge as $F \rightarrow 0$, however $T_F$ and $T_G$ are finite and independent of $F$ \cite{srebro_levine}, as is the case when $x$ is height and $F$ a gravitational field.
\bibitem{2nd_moms}
The procedure used in \cite{srebro_levine} to calculate all moments of $e_i$ has been used to obtain the steady state averaged second moments, which for $N \gg 1$ reduce to \\
%
$\langle x_i^2 \rangle - \langle x \rangle ^2 = f(1-h) T_B^2 / F^2 [f^2(h^2+2h-1) + 2(f(1-3h)+A_1)A_1 + f(1-h)(3-A_2)] / [(2A_2-fh)(A_1-fh)^2]$,
%
\\
%
$\langle X E \rangle  - \langle X \rangle \langle E \rangle = - N T_B^2/F \cdot f^2h(1-h) [4A_1^2(3A_1+3+f(2-9h)-A_2) +2fA_1(15-27h-f(5-12h-3h^2)-(5-9h)A_2) -f(18(A_2-1)(1-h)+f(12-21h-3h^2)-f(A_2+f)(2-3h-3h^2))] /[6(A_1-fh)^3(A_1+fh)(2A_2-fh)]$,
%
\\
%
$\langle X^2 \rangle - \langle X \rangle^2 = N [ \langle x_i^2 \rangle - \langle x_i \rangle ^2 ] - [ \langle XE \rangle - \langle X \rangle \langle E \rangle ] /F,
$
%
\\
%
with $A_n \equiv n + (1-f)(1-2 \alpha^n)$.
\bibitem{baldassarri_footnote}
The Einstein relations, derived here from the Kubo formula with $\omega=0$, were previously shown to hold for the 1D Maxwell model with a Gaussian thermostat \cite{baldassarri_2005}.
\end{thebibliography}
\end{document}